# Photometric properties of new solar H$\alpha$ commercial Fabry-Perot etalons.
# Application to the analysis of the chromospheric fringe.

by

Cyril Bazin and Serge Koutchmy

*Institut d'Astrophysique de Paris, UMR 7095, CNRS and UPMC*
*98 Bd Arago F-75014 Paris (France)*
*bazin@iap.fr    koutchmy@iap.fr*

## Abstract

We consider the use of the commercially available Fabry-Perot etalons (FP) of Sidorin and Lunt type for the imaging of the solar chromosphere in the H$\alpha$ line of HI. Three etalons of 40, 60 and 90 mm diameter were evaluated and accurately analysed. At normal incidence the maximum transmission wavelength was 656,285 nm for the 60 and 40 mm FP etalons. The finesse has been evaluated at 13,3 for the FP 60mm, 8,7 for the FP 40 and 13,9 for the FP 90 mm. Shifts of the central wavelength of maximum transmission were accurately measured as a function of the incidence angle. Polynomial curves precisely fitting the transmitted central wavelength variations when using a quasi-parallel beam from a point-like source are presented. Further calibrations have been done with photometric accuracy using a laboratory set-up comprising i/ a 16 bits CCD camera; ii/ a Littrow spectrograph of a spectral power 110000 giving a linear dispersion of .0058 nm/pixel on the CCD and iii/ an "artificial Sun" used as a light source and iv/ precisely adjustable in position optical components, including the F-P etalons. In addition, a precise laboratory wavelength calibration was performed using a low pressure deuterium $^2$D spectral lamp simultaneously illuminating the adjustable entrance slit using a splitter before. The variations of the FWHM of the spectral transmission variations as a function of the incidence angle of a parallel beam are also given for each etalon. Consequences resulting from the use of a low but significant aperture/ratio are tentatively discussed for the first time. An application to a precise photometric work of solar physics interest when using limb filtergrams is illustrated and discussed, with emphasis on the photometric accuracy resulting from the use of such etalons put before the entrance aperture of an imaging telescope. Monochromatic images of the solar chromosphere shell in the vicinity of the polar and equatorial limbs were made using a small telescope, in order to deduce the variation of the typical average thicknesses at poles and equator interpreted as a prolateness effect of the chromospheric shell observed during the last minimum of solar activity (2009).

## Content

# I) Introduction

Images of the entire solar disk obtained with a Fabry Perot (FP) etalon are suffering from the problem of the transmission wavelength *shift* especially important in the case of solar works where a minimum photometric accuracy is required [1] [2]. Accordingly, the transmitted intensities can change over the field of view [3]. In order to illustrate and give practical advice, we use FP of the Coronado Company which produces such etalons without giving any high spectral resolution calibration curves of the FP they sell [4] [5] [6] [7], to perform laboratory calibration with a photometric accuracy.

This work aims at experimentally measuring the Full Width at Half Maximum (FWHM) and the wavelength shift of the transmitted light corresponding to H$\alpha$ line of HI when the FP is used for solar measurements before the entrance aperture of the imager. This type of observation is also broadly used by amateurs and can be of interest to cover the chromospheric activity. After calibrations, the FP filters were tried indeed on the chromosphere [8] to study the shell [20]. First, we characterized the central wavelenght shift using an optical setup composed of a collimator illuminating the FP with a quasi parallel beam (< 4 arc min). The transmitted light through the FP is then analysed using a Littrow spectrograph with a resolving power > 110000.

## II) Optical Setup used to perform the spectro-photometry of the etalons

We design and built the following optical setup to analyse the spectra through the FP.

It is a Littrow spectrograph design [9].

We used an artificial source of white-light W-L with a diaphragmed tungsten arc lamp that produces a beam less than 4 arcmin of aperture ratio that illuminates the FP surface. The diagram of Fig.1 describes the set-up. Note that the diaphragm 2 is adjustable up to a value corresponding to an extended source of 32' matching the angular size of the Sun in the sky. The source is then called an "artificial" Sun.

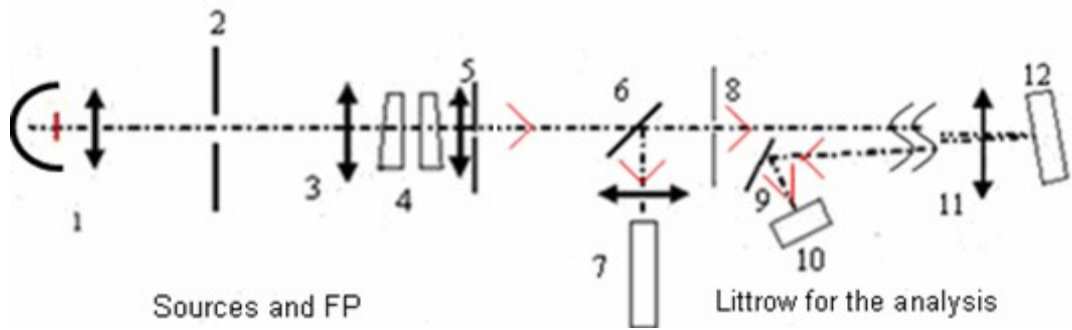

**Figure 1** Optical setup made of : 1) & 2) artificial source part; 3) collimating lens, F = 300 mm; 4) FP etalon; 5) 600 mm focal lens reimaging the diaphragm 2 on the entrance slit 8; 6) beam splitter; 7) low pressure calibration deuterium lamp; 8) entrance adjustable slit; 9) plane mirror; 10) 16 bits CCD camera; 11) 1560 mm focus Littrow lens; 12) plane reflective grating 100x140 mm $^2$

A low pressure deuterium $^2$D arc lamp is introduced in the setup (see Fig 1), in order to produce the D$\alpha$ line at 656.12 nm that is closed to the H$\alpha$ line for wavelength calibrations [10]. This allows having an emission line taken as reference for wavelength in the same field of view of the CCD 16 bit camera. The FP etalon is mounted in rigid fixations in the optical setup that allows to make the measurements easily, see figures 2, 3 and 4.

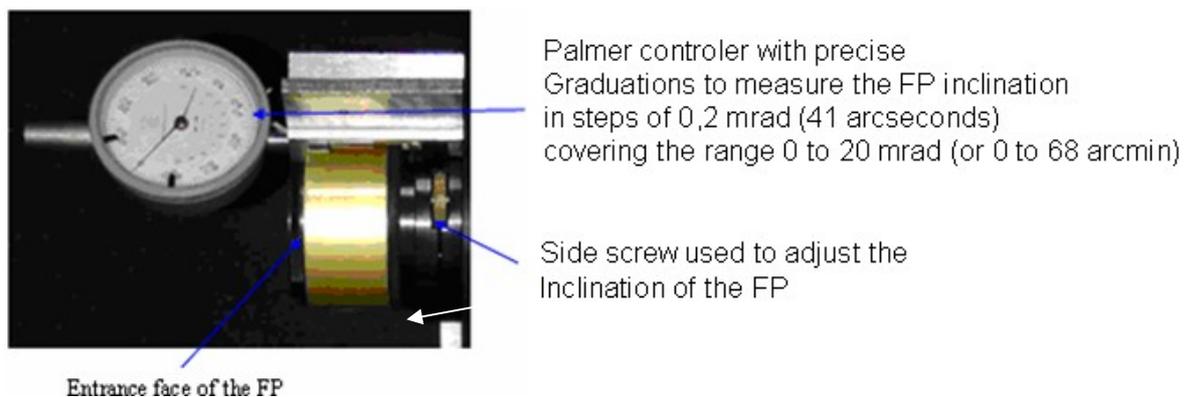

**Figure 2** : View of the micrometer used for the inclination angle measurement of this FP

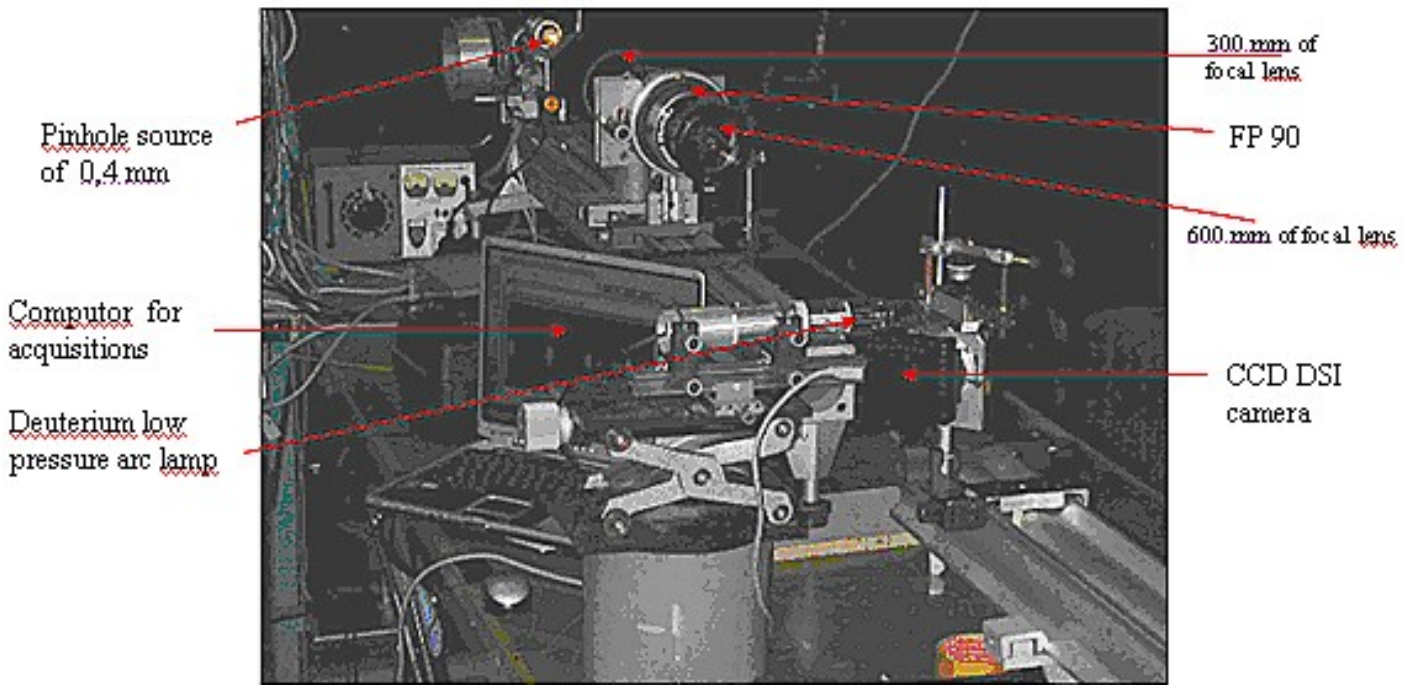

**Figure 3**: View of the setup when using a calibrated pinhole arc lamp of 0.4 mm aperture as a source instead of the extended artificial Sun

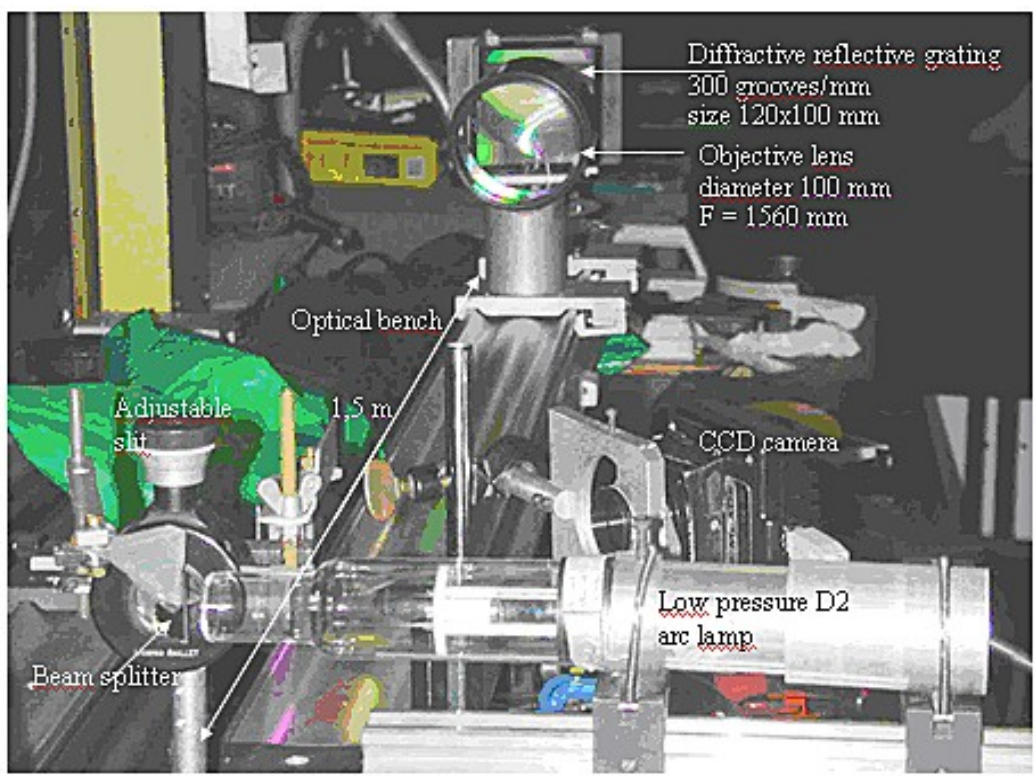

**Figure 4** : View of the spectrograph from the entrance slit to show the set-up with the calibration $^2$D lamp.

## III) Data acquisition and analysis

The acquisitions were made using a 16 bit DSI CCD Meade Camera, connected to a computer. The Autostar Meade software is used for the acquisitions. A blocking narrow pass band filter was inserted in order to select the main fringe (in French: "cannelure") near 656.28 nm. This blocking filter had a peak of transmission near H$\alpha$ and a width FWHM of 1.6 nm. This explains why the 644$^{th}$ order ring is seen with a much lower intensity, in the field of view of the CCD, see Fig. 5.
The D$\alpha$ line is close to the bright 643$^{rd}$ order fringe, which corresponds to the transmission of the H$\alpha$ line.
The finesse can be evaluated, see IV). Each measurement is made with the FP 60 inclination angle, starting with the normal incidence to a maximal inclination of 0.02 rad by a 0.0002 rad step. We compared also the results obtained with a FP 40 mm and FP 90 mm diameter.
The following graphs and pictures show the results obtained at room temperature:

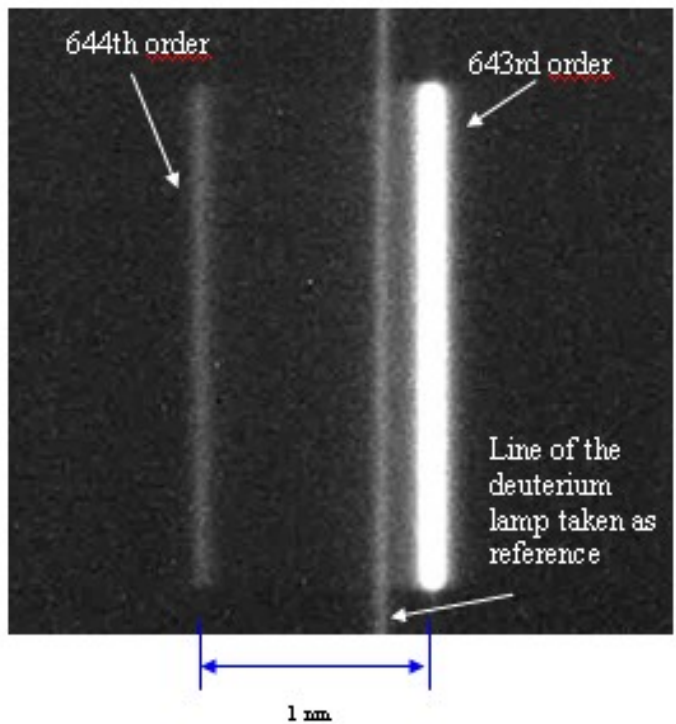

**Figure 5:** Typical calibration image showing the fringes of the FP illuminated by the W-L source and by the lamp producing the D$\alpha$ line. Note that a blocking filter is used to make this image, allowing the 643rd order to be well transmitted.

The image in figure 5 is showing the transmitted spectrum of the fringe of the FP 60, at normal incidence. The reference $D\alpha$ line is in the same field of view and close to the 643$^{rd}$ order. The $D\alpha$ line gives also the instrumental response because the narrow line profile corresponds to a very low pressure gaz [11] [12] in the lamp, and the instrumental FWHM is evaluated at 0.045 nm.

In the centre of the image, the line of the $D\alpha$ lamp is fixed and the FP fringe moves towards the blue wavelength direction when the tilt angle of the FP increases.

The use of etalon FP placed before a telescope is possible and is described by the manufacturer Coronado [14]. But images are suffering of the wavelength shift in the blue wing of $H\alpha$ and also by the enlargement of the FWHM that is dependent of the aperture ratio of the instrument. These spectro-photometrical measurements allow to determine the effective finesse [15] [16], and the transmitted wavelength with the inclination angle of the beams on the FP are given in the following graphs, figures 6 to 8. Additional calibrations curves for the other FP are given in the Annexes, figure 1 to 3.

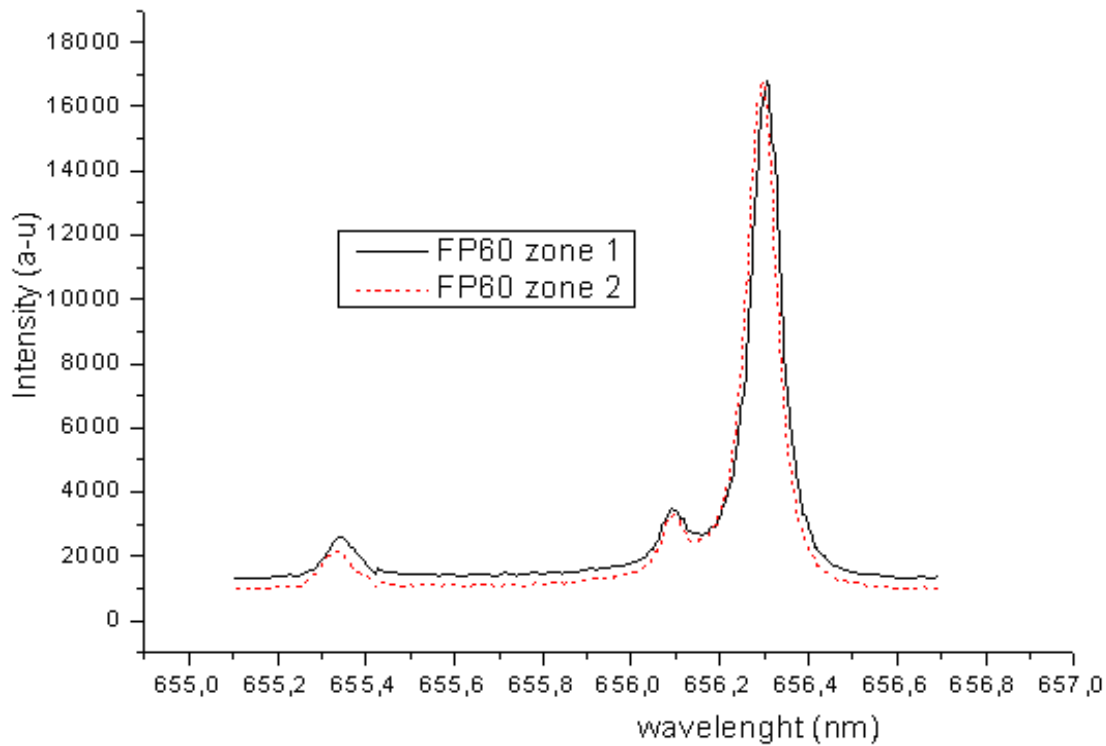

**Figure 6:** Graph illustrating the intensity profiles of the transmission spectrum using the $^2D$ lamp.

The two profiles show the spectrum of the transmitted light taken in two parts (zone 1 and 2) of the FP surface, in order to evaluate the quality of the surfaces of the FP mirrors.
This spectrum allows also evaluating a dispersion of 0.006 nm/pixel or appr. 0.6 nm/mm

## IV) Results

The effective finesse, the resolving power were measured for the FP 60. The FP40 and 90 mm are also compared. In the setup, we replaced the FP 60 by the FP 40 and FP 90. We measured the transmitted wavelength with the solar spectrum, in order to see the wavelength shift in the blue H$\alpha$ wings. The following table and graphs show those spectral parameters near the wavelength of H$\alpha$.

**Finesse F measurements:**

| Type and diameter of the FP | Distance between two Consecutive cannelures in nm | FWHM of the cannelure 643 corresponding with H$\alpha$ in nm | Resolving power at 656,285 nm |
|---|---|---|---|
| FP 40 mm | 0,79 | 0,091 | 7211 |
| FP 60 mm | 0,957 | 0,072 | 9115 |
| FP 90 mm | 1,073 | 0,076 | 8552 |

$$F = \frac{0,957}{0,072} = 13,3 \pm 0,5 \text{ for the FP de 60}$$

$$F = \frac{0,79}{0,091} = 8,7 \pm 0,5 \text{ for the FP de 40 and } F = \frac{1,073}{0,076} = 13,9 \pm 0,5 \text{ for the FP 90}$$

**Table 1:** parameters of the FP deduced from the intensity profiles of the FP spectrum

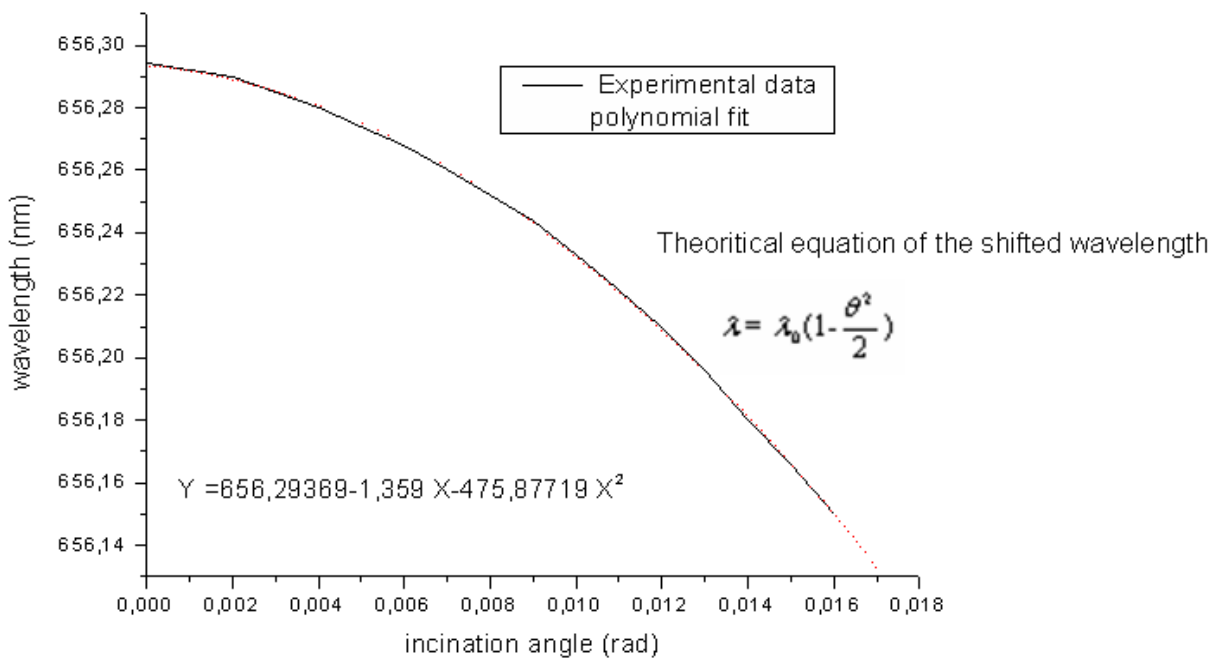

**Figure 7:** Graph giving the wavelength shift with the inclination angle of the Fabry-Perot

The wavelength variation of the maximum transmission wavelength with the inclination angle is polynomial, and in good agreement with the theory of the FP.

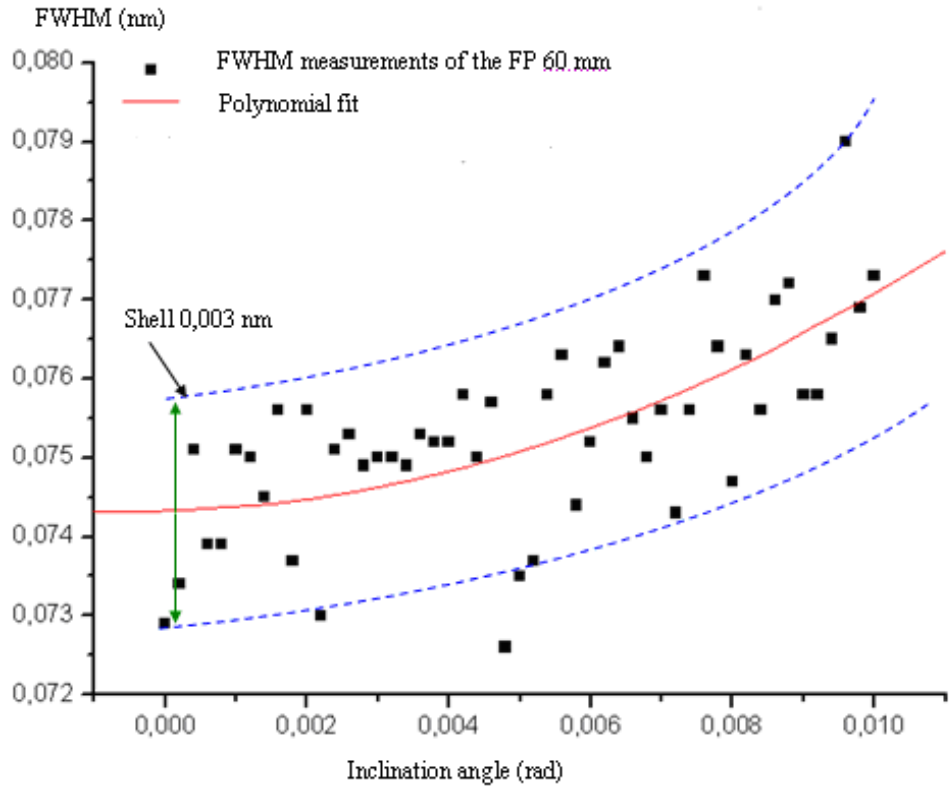

**Figure 8:** Graph illustrating the FWHM shift with the inclination angle of the FP

The FWHM of the 643$^{rd}$ order fringe increases with the inclination angle, and the shape is also polynomial, in good agreement with the theory, and the equation 1, given in part V)

## V) Study of the FWHM H$\alpha$ transmission curves using different aperture ratio

The FWHM measurement with different aperture ratio was performed by changing the diameter of the diaphragm placed in front of the objective N°5 in the diagram figure 1.
This component is placed before a 600 mm focal lens used to focus the image on the entrance slit. The beams were also oriented in normal incidence regarding the entrance of the FP.
An acquisition was made for every diameter of the diaphragm. For this measurement, we changed only the diameter of this diaphragm without moving it, and made an acquisition at every aperture.

The draw back of the increase of the FWHM of the FP fringe while the aperture ratio decreases, is that the filter becomes less selective in the H$\alpha$ line, more light is coming from the tails of the line, and the images of the full disc of the Sun have a lower contrast.
The following curve in Figure 9 shows the FWHM variations corresponding to H$\alpha$, given by the FP 60 with different aperture ratio (F/D):

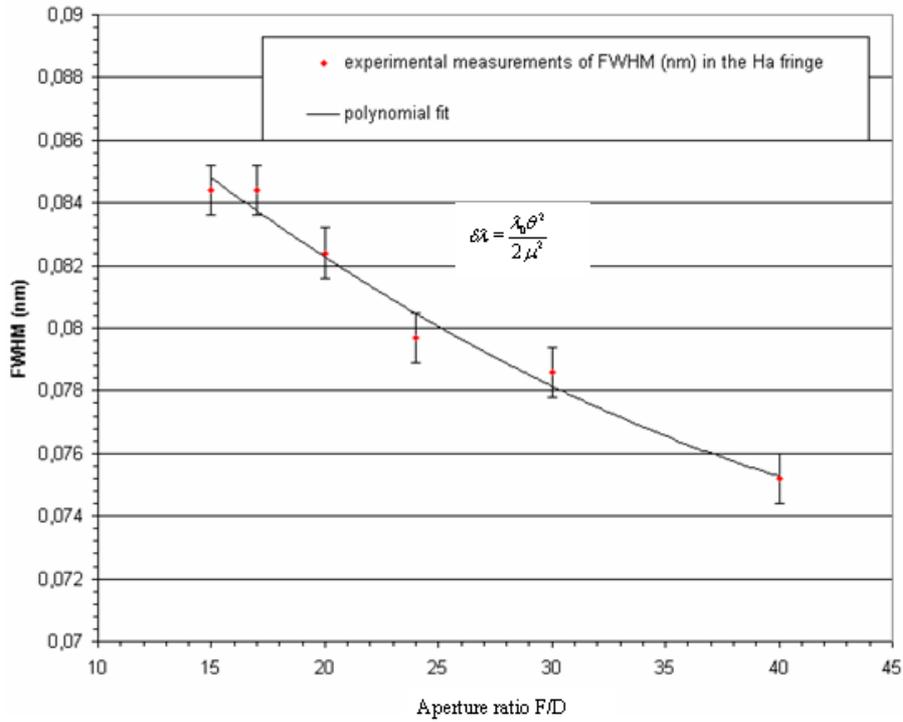

**Figure 9:** FWHM shift as a function of the aperture ratio for the FP of diameter 60 mm

The FWHM of the fringe increases when the aperture ratio decreases.
The equation that gives the FWHM with the inclination angle is:

$$\delta\lambda = \frac{\lambda_0 \theta^2}{2\mu^2} \quad \text{(equation 1) [17]}$$

$$\tan\theta = \frac{D}{F} = \frac{1}{AN} \# \theta$$

$\lambda_0$ is the wavelength of H$\alpha$ 656.285 nm and µ is the effective refractive index of the dielectric layers

By using this equation fit, we can deduce other parameters such as the refractive indices of the layers on the mirrors of the FP:
The following table gives the calculated indices µ corresponding to experimental results:

| aperture (F/D) | $\theta$ (°) | refractive index µ |
|---|---|---|
| 40 | 1,432391138 | 1,651438213 |
| 30 | 1,909854851 | 2,153766976 |
| 24 | 2,387318564 | 2,673565564 |
| 20 | 2,864782277 | 3,155278075 |
| 17 | 3,37033209 | 3,667846031 |
| 15 | 3,819709702 | 4,156892169 |

**Table 2:** indices µ corresponding to our experimental results

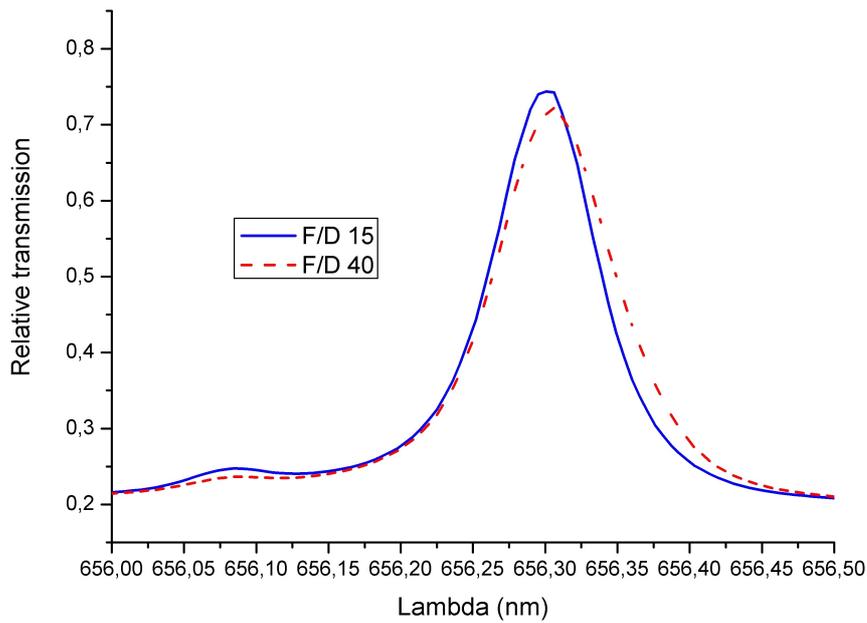

Figure 10: transmitted wavelength near H$\alpha$ with different aperture ratio

The F/D = 15 aperture ratio seems to be shifted by 0.1 Angstrom to the blue wavelength compared to the case of F/D= 40.
Also we analysed the transmission wavelengths of the filters when using the solar spectrum, as the Figure 11 shows.

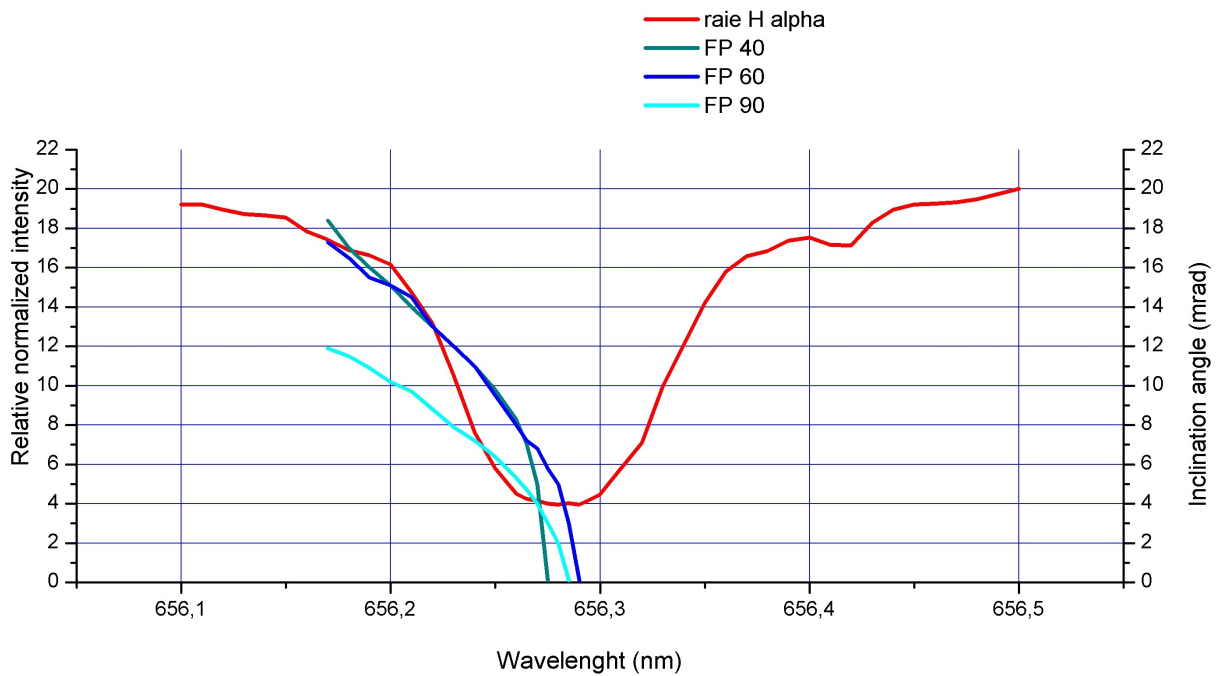

**Figure 11:** Central wavelength variations of the effective transmission. 3 studied FP when the inclination angle in mrad is changed and the reproduction of the solar spectrum near H$\alpha$.

The resulting indices values increase more rapidly because the 2 mirrors of the FP Coronado filters are coated. The company doesn't give any value about the effective refractive indices of the layers of the FP filters they sell. The graphs in fig 10 give the central and maximal transmission wavelength shift with the aperture ratio. This was accurately evaluated. The graph in figure 11 illustrates the wavelength variations of the 3 FP compared with the inclination angle, with the tilt angle and fitting with the H$\alpha$ line profile.

These filters at normal incidences don't give the same wavelength transmission value when they are compared to each other, and the graphs show that for a bigger diameter of a Coronado FP, the wavelength shift increases more rapidly with the inclination angle, than for a smaller diameter of the FP.

Knowing these parameters, it was then possible to use them to try measurements, in small fields of view, of the chromosphere thickness.

## VI) Chromosphere thickness and "ovalisation" measurements

A first analysis of the chromosphere shell was performed by making a superposition of the solar limb spectrum and the FP 643$^{rd}$ fringe in order to measure its position in the H$\alpha$ line.

Figure 12 shows a spectrum of the solar limb focussed at the entrance slit of a small solar spectrograph, and the FP 60 placed in front of the slit in order to simultaneously obtain over the same image the spectrum of the solar spectrum in Ha in different parts of the limb by cutting the shell of H$\alpha$ emission. Some other absorption lines are seen in the spectrum and can be identified thanks to the spectral atlas [13]. This technique allowed trying to observe the chromosphere in the H$\alpha$ line and in the wings, and trying to make velocities measurements in the shell.

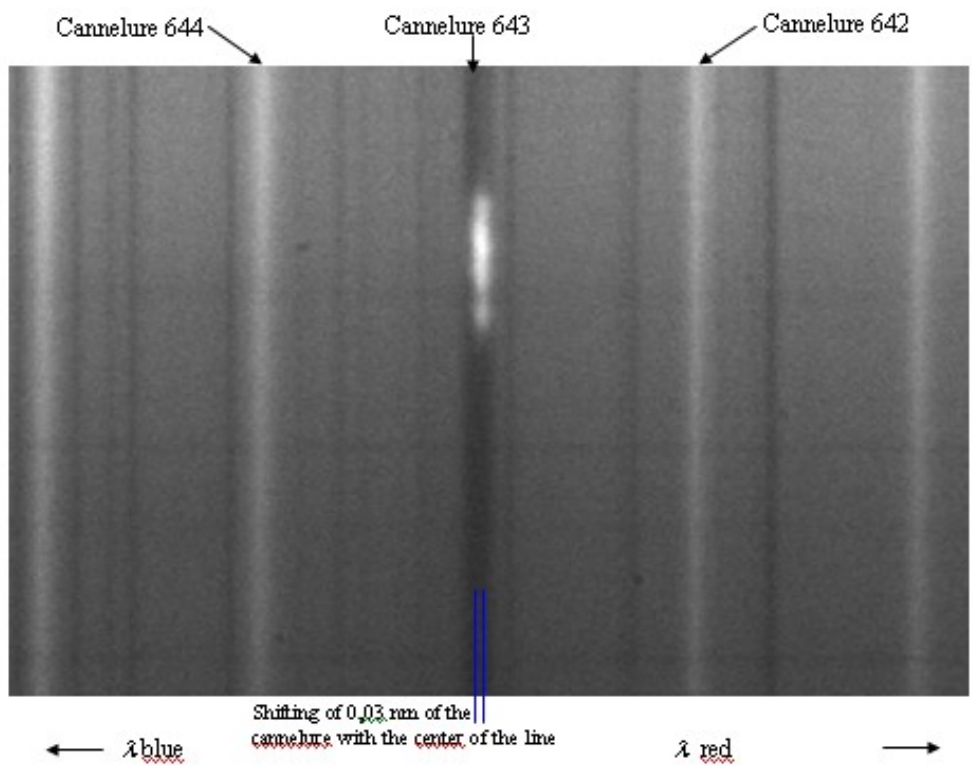

**Figure 12:** image showing the 2D spectrum given by the FP 60 mm with the slit at the solar limb near H$\alpha$

Here we call ovalisation the prolatness, which means a solar diameter longer in the North-South direction.

We determined the spectral parameters of the FP and used them for the chromosphere prolateness studies in $H\alpha$ [18], and we made images with the calibrated etalon FP filters in a field of view of 5x5 arcmin of the chromosphere. This allows keeping a constant wavelength transmission in all the field of view.

The FP 90 mm was placed in front a refractive fluorite lens of 1500 mm of focal length. Some pictures of each solar limb were made on 18[th] March 2009 with a clear sky.

These images are presented in the Annexe, figure N°6 .

A 5 minutes acquisition sequence of images was taken at North Pole, then just after at Equators limbs and at South Pole. The scale was 1 pixel for 562 km.

A summation of 80 lines of pixels on the reduced images was done, with neglecting the limb curvature. This allows increasing the signal to noise ratio and to see the shape of the chromosphere intensity profile in $H\alpha$ .

Then we plotted and adjusted the intensity profiles of the North pole and West equator limbs, and we obtained the following graphs:

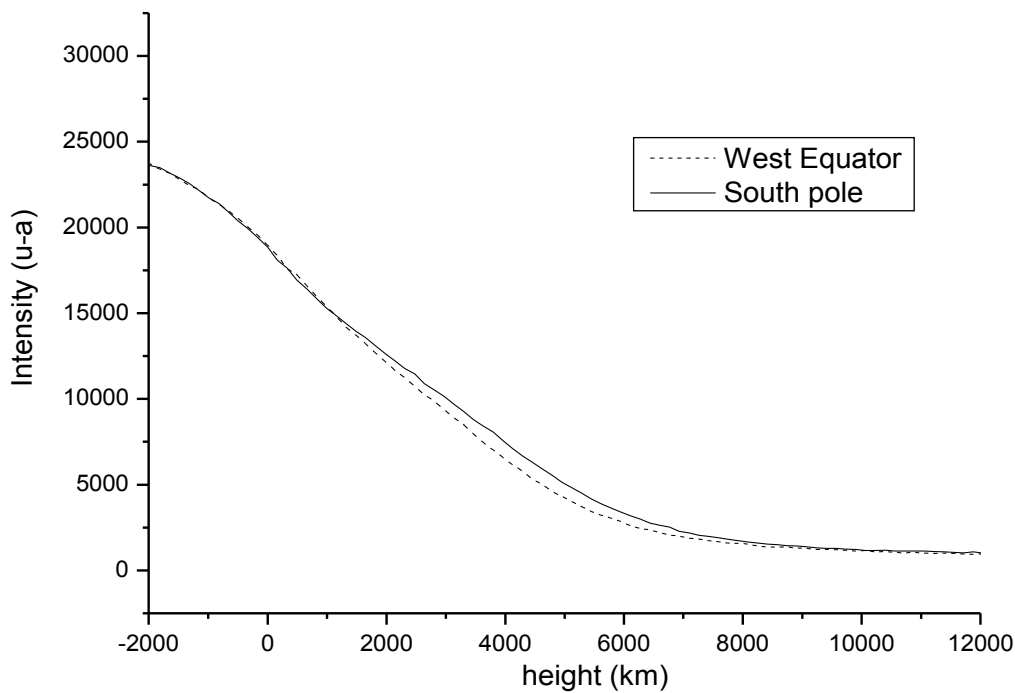

**Figure 12:** intensity profiles after integrating over 80 lines of pixels, at equators and poles

The thickness of the chromosphere is measured to be 6000 km and a difference of height about 500 +/-300 km is identified between the North Pole and Equator. The North Pole is higher than the Equator, and it was observed in a low turbulence day, 18th Mars 2009 with a clear sky, at this period of low activity of the Sun. These measurements have also been done in 2005 and 2006 in the solar minimum [19] and the new results confirm the ovalisation effect.

# VII) Conclusion:

The setup allowed us to obtain reproducible results with a hysteresis lower than 0.002 nm. A finesse of the FP up to 14 is used for a limited field of view 5'x5' where the wavelength shift is small. The FWHM of the FP is evaluated at 0.075 nm for a F/D = 40, close to the nominal value, and increases to 0.085 for F/D = 15. This shows the photometric limitations of the FP Coronado.

The thickness of the chromosphere is measured to be about 6000 km and the ovalisation effect to 500 +/- 300 km in the $H\alpha$ line studied with this etalon in 2009.

We plan to add polarisers or crystals in the beam path of the setup in order to obtain double pictures with the same intensity, but in orthogonal polarisations over the same CCD picture, and by subtraction we expect to build high resolution magnetograms. This technique could allow reducing the turbulence effects in the polarized pictures during the 10 frames/s acquisition rate thanks to the new generation of fast CCD 12 bit Lumenera cameras.

**ANNEXES : Additionnal graphs, diagrams and images**

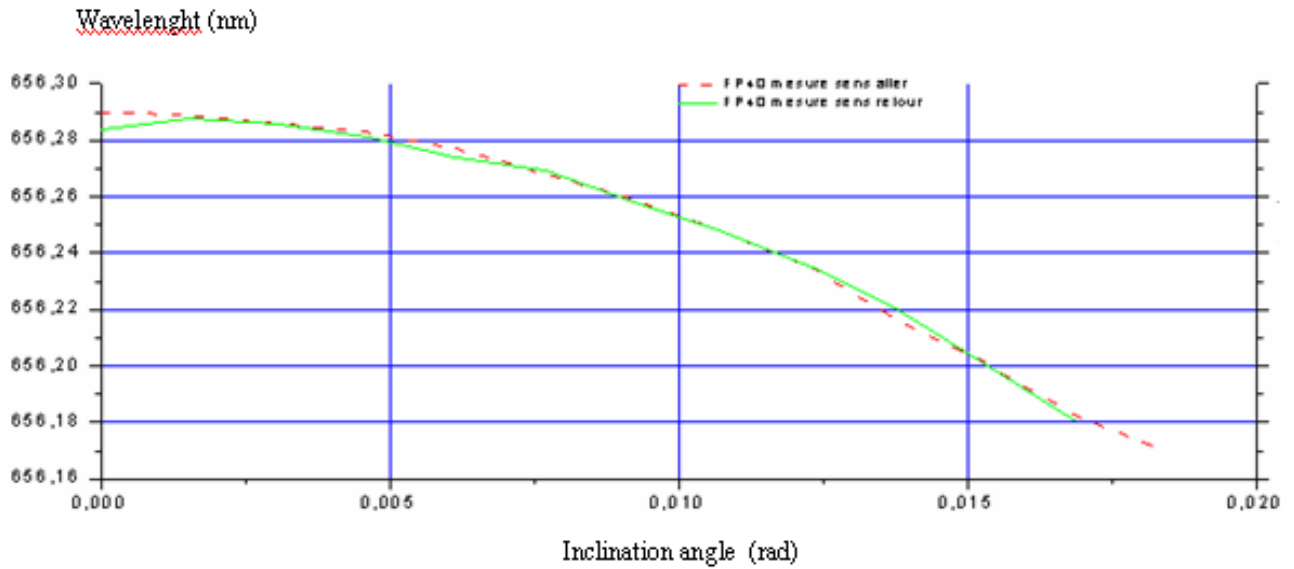

**Annexe 1:** Transmitted wavelength measurement in function of the inclination angle of the FP 40 for the 643$^{th}$ fringe corresponding to the H$\alpha$ line

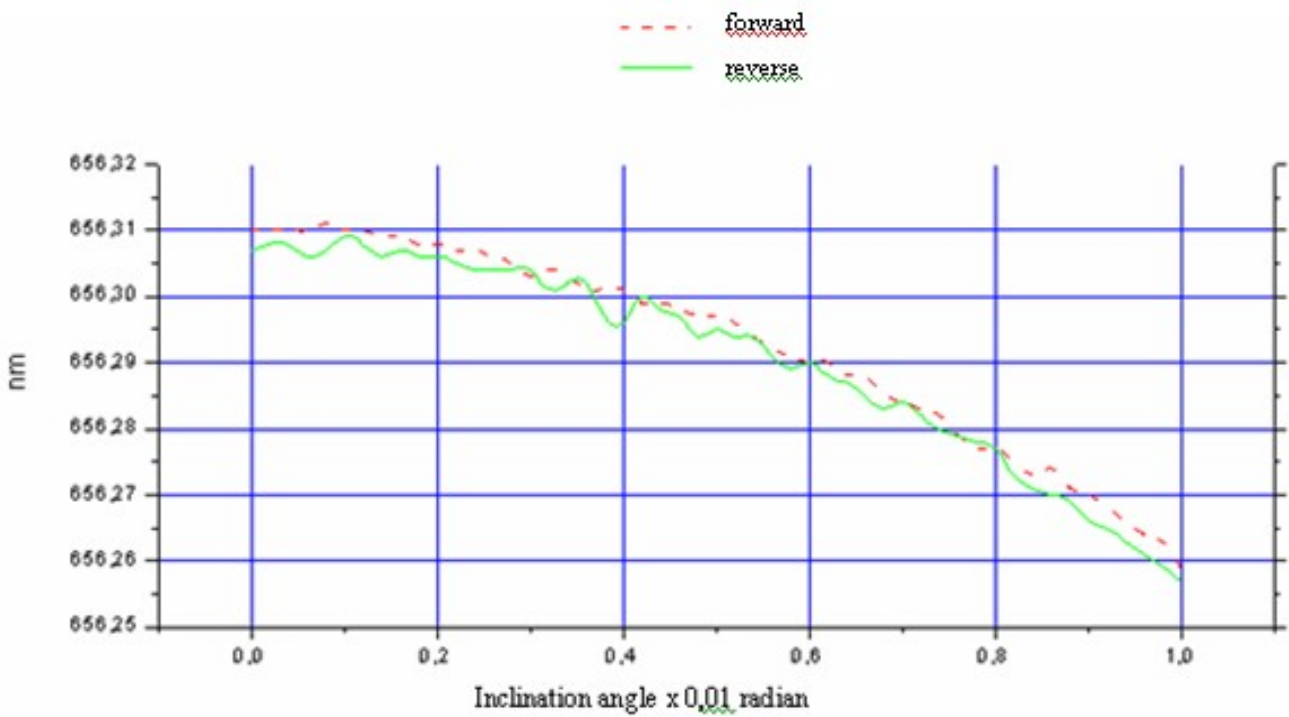

**Annexe 2:** Shift of the H$\alpha$ 643$^{th}$ fringe with the inclination angle of the FP 60 mm

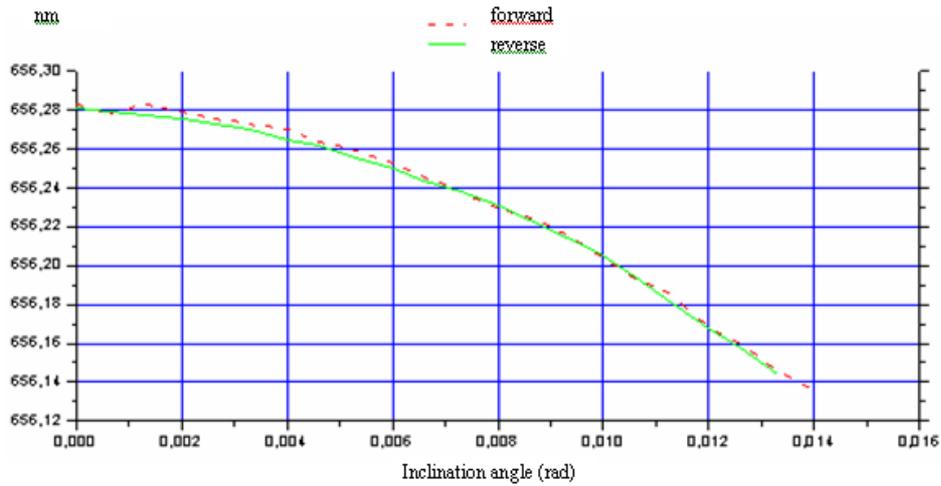

**Annexe 3:** Transmitted wavelength measurement with the inclination angle of the FP 90 for the 643$^{th}$ fringe

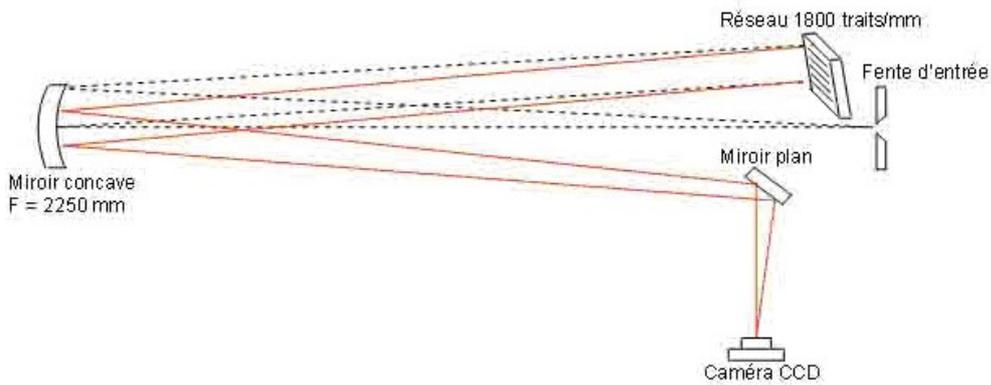

**Annexe 4:** Diagram of the spectrograph used to obtain the spectrum of figure 6

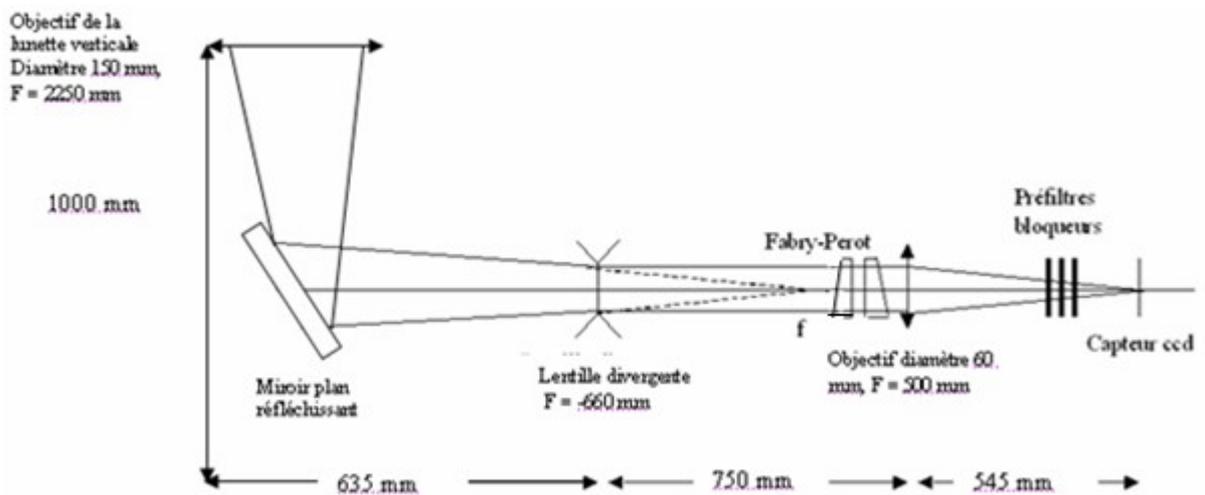

**Annexe 5:** Optical diagram of the telecentric optical setup

Images used for the solar chromosphere measurements:

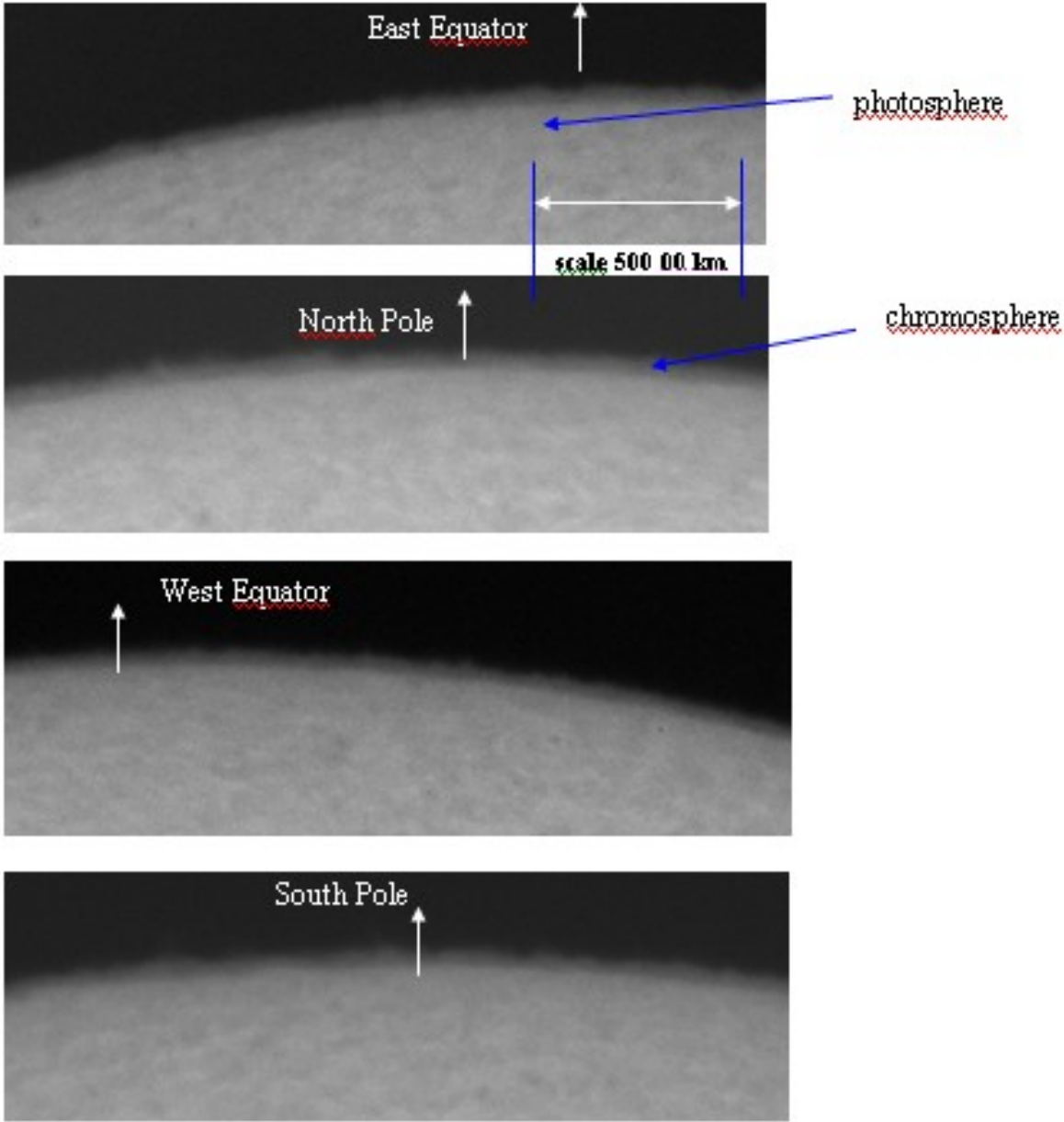

**Annexe 6:** Images of the solar chromosphere taken on March 18th 2009 by Sylvain Weiller in Saint Remy les Chevreuse, south of Paris, with a 150/1500 mm Fluorite refractor fitted with the FP 90 mm Coronado.

16